\documentstyle[12pt]{article}

\newfont{\DamirFont}{cmr10}
\renewcommand{\footnote}[1]{%
\def\thefootnote{\arabic{footnote})}
\footnotemark%
\footnotetext{#1}}
\newfont{\blackboard}{cmr10}
\newcommand{\Z}{\mbox{\blackboard\symbol{"5A}}}

\newcommand{\io}{[\hspace{-1pt}[}
\newcommand{\ic}{]\hspace{-1pt}]}
\newcommand{\fo}{\{\!\mid\!}
\newcommand{\fc}{\!\mid\!\}}

\newcommand{\sgn}{{\rm sgn}}

\renewcommand{\Re}{{\rm Re\,}}
\newcommand{\REG}{{\rm REG}}
\newcommand{\IRREG}{{\rm IRREG}}

\newcommand{\tr}{{\rm tr}}

\renewcommand{\arctan}{{\rm arctan}}

\renewcommand{\cosh}{{\rm cosh}}

\newcommand{\Lp}{L_{(+)}}
\newcommand{\Lm}{L_{(-)}}

\def\be{\begin{equation}}
\def\ee{\end{equation}}
\def\bea{\begin{eqnarray}}
\def\eea{\end{eqnarray}}
\def\ba{\begin{array}}
\def\ea{\end{array}}
\def\V{{\bf V}}
\def\J{{\bf J}}
\def\j{{\bf j}}

\def\vac{{\rm vac}}
\def\x{{\bf x}}
\def\F{\Phi^{(0)}}
\def\M{{\cal M}}
\def\N{{\cal N}}
\def\S{{\cal S}}
\def\L{{\cal L}}
\def\A{{\cal A}}
\def\C{{\cal C}}

\def\i{{\int}}

\newcommand{\bpar}{\mbox{\boldmath $\partial$}}
\newcommand{\ab}{\mbox{\boldmath $\alpha$}}
\newcommand{\gb}{\mbox{\boldmath $\gamma$}}

\def\si{\mathop{\displaystyle\sum\mkern-22mu\int\,}}

\newcommand{\ds}{\displaystyle}
\newcommand{\sss}{\scriptscriptstyle}

\begin{document}

\begin{center}
{\bf POLARIZATION OF THE MASSLESS FERMIONIC VACUUM IN THE BACKGROUND OF
A SINGULAR MAGNETIC VORTEX IN 2+1-DIMENSIONAL SPACE-TIME}

\vspace{0.5cm}
(Ukrainian Journal of Physics, {\bf 43}, no.12, 1513-1525 (1998))

\vspace{0.5cm}
{\bf YU.A. SITENKO}

\vspace{0.5cm}
Bogolyubov Institute for Theoretical Physics, National Academy of
Sciences,\\ 14b, Metrologichna str., Kyiv 252143, Ukraine
\end{center}

\begin{abstract}
Effects of the configuration of an external static magnetic field in
the form of a singular vortex on the vacuum of a quantized massless
spinor field are determined. The most general boundary conditions at
the punctured singular point which make the twodimensional Dirac
Hamiltonian to be self-adjoint are employed.
\end{abstract}

\section{Introduction}

A study of effects of singular external fields (zero-range potentials)
in quantum mechanics has a long history and has been comprehensively
conducted (see \cite{Alb} and references therein). Contrary to this,
effects of singular external fields in quantum field theory are at the
initial stage of consideration, and much has to be elucidated. Singular
background can act on the vacuum of a second-quantized spinor field in a
rather unusual manner: the leakage of quantum numbers from the
singularity point occurs. This is due, apparently, to the fact that a
solution to the Dirac equation, unlike that to the Klein-Gordon one,
does not obey a condition of regularity at the singularity point. It is
necessary then to specify a boundary condition at this point, and the
least restrictive, but physically acceptable, condition is such that
guarantees self-adjointness of the Dirac Hamiltonian. Thus, effects of
polarization of the vacuum by a singular background appear to
depend on the choice of the boundary condition at the singularity
point, and the set of permissible boundary conditions is labelled, most
generally, by a self-adjoint extension parameter.

As examples of singular background configurations, one can consider
a pointlike magnetic monopole in threedimensional space and a pointlike
magnetic vortex in twodimensional space. While in the first case there
is a leakage of charge to the vacuum, which results in the monopole
becoming the dyon violating the Dirac quantization condition and CP
symmetry \cite{Wit,Gro,Yam}, in the second case the situation is much
more complicated, since there is a leakage of both charge and other
quantum numbers to the vacuum. For particular choices of the boundary
condition at the singularity point it has been shown that charge
\cite{Sit88,Sit90}, current \cite{Gor} and angular momentum
\cite{SitR96} are induced in the vacuum. The vacuum quantum
numbers under general boundary conditions which are  compatible with
self-adjointness have been considered in Refs. \cite{Sit96, Sit97,
SitR97}.

Thus far the effects of polarization of the massive fermionic vacuum
have been studied. In an irreducible representation of the Clifford
algebra in 2+1-dimensional space-time, the mass term violates the symmetry
under both space and time parity transformations. Our interest will be
in the parity of the third type, which is similar to the axial symmetry
in evendimensional space-times (see Refs.\cite{Temp,Alva1, Alva2}).
This parity is also violated by the mass term in the irreducible
representation.

In the absence of the mass term all above symmetries are formally
present. Thus our concern is, in particular, whether the polarization 
of the massless vacuum in a singular background respect these formal
symmetries?
It will be shown that, although the parity anomaly  is absent, the
parity breaking condensate emerges  in the vacuum. In a
parity-invariant model with two species of fermions  composing a
reducible representation of the Clifford algebra, this condensate is
transformed into the chiral symmetry breaking condensate. Also all
other characteristics  of the massless fermionic vacuum in a singular
background are determined.

\section{Quantization of Spinor Field and Boundary Condition at the
Location of the Vortex}

The operator of the second-quantized spinor field is presented in the
form
$$
\Psi(\x,t)=\si_{E_\lambda>0}\, e^{-iE_\lambda
t}<\x|\lambda>a_\lambda+\si_{E_\lambda<0}\, e^{-iE_\lambda
t}<\x|\lambda>b_\lambda^+, \eqno(2.1)
$$
where $a_\lambda^+$ and $a_\lambda$ ($b_\lambda^+$ and $b_\lambda$) are
the spinor particle (antiparticle) creation and annihilation operators
satisfying the anticommutation relations
$$
[a_\lambda,a_{\lambda'}^+]_+=[b_\lambda,b_{\lambda'}^+]_+=<\lambda|\lambda'>,
\eqno(2.2)
$$
and $<\x|\lambda>$ is the solution to the stationary Dirac equation
$$ H<\x|\lambda>=E_\lambda<\x|\lambda>, \eqno(2.3)$$
$H$ is the Dirac Hamiltonian, $\lambda$ is the set of parameters
(quantum numbers) specifying a state, $E_\lambda$ is the energy of
a state; symbol $\si$ means the summation over discrete and the
integration (with a certain measure) over continuous values of
$\lambda$. The ground state $|\vac>$ is defined conventionally by the
equality
$$ a_\lambda|\vac>=b_\lambda|\vac>=0. \eqno(2.4) $$
In the case of quantization of a massless spinor field in the
background of a static vector field $\V(\x)$, the Dirac Hamiltonian
takes the form
$$ H=-i\ab[\bpar-i\V(\x)], \eqno(2.5)$$
where
$$\ab=\gamma^0\gb, \qquad \beta=\gamma^0, \eqno(2.6)$$
$\gb$ and $\gamma^0$ are the Dirac $\gamma$ matrices. In the
2+1-dimensional space-time $(\x,t)=(x^1,x^2,t)$, the Clifford algebra
has two inequivalent irreducible representations which can be differed
in the following way:
$$i\gamma^0\gamma^1\gamma^2=s, \qquad s=\pm1.
\eqno(2.7)$$
Choosing the $\gamma^0$ matrix in the diagonal form
$$\gamma^0=\sigma_3, \eqno(2.8)$$
one gets
$$\gamma^1=e^{{i\over2}\sigma_3\chi_s}i\sigma_1e^{-{i\over2}\sigma_3\chi_s},
\quad \gamma^2=
e^{{i\over2}\sigma_3\chi_s}is\sigma_2e^{-{i\over2}\sigma_3\chi_s},
\eqno(2.9)$$
where $\sigma_1,\sigma_2$ and $\sigma_3$ are the Pauli matrices, 
and $\chi_1$ and
$\chi_{-1}$ are the parameters that are varied in the interval
$0\leq\chi_s<2\pi$ to go over to the equivalent representations.

The configuration of the external field $\V(\x)=(V_1(\x),V_2(\x))$ is
chosen to be
$$V_1(\x)=-\Phi^{(0)}{x^2\over(x^1)^2+(x^2)^2}, \quad V_2(\x)=\F
{x^1\over(x^1)^2+(x^2)^2}, \eqno(2.10)$$
which corresponds the magnetic field strength in the form of a singular
vortex
$$\bpar\times\V(\x)=2\pi\F\delta(\x),
\eqno(2.11) $$
where $\F$ is the total flux (in $2\pi$ units) of the vortex -- i.e.,
of the thread that pierces the plane $(x^1,x^2)$ at the origin. The
wave function on the plane with the punctured singular point $\x=0$
obeys the most general condition (see \cite{Sit96} for more details)
$$<r,\varphi+2\pi|=e^{i2\pi\Upsilon}<r,\varphi|, \eqno(2.12)$$
where $r=\sqrt{(x^1)^2+(x^2)^2}$ and $\varphi=\arctan(x^2/x^1)$ are the
polar coordinates, and $\Upsilon$ is a continuous real parameter which
varies in the range $0\leq\Upsilon<1$. It can be shown (see, for
example, \cite{SitR96,Sit96}) that $\Upsilon$ as well as $\F$ is
changed under singular gauge transformations, whereas the difference
$\F-\Upsilon$ remains invariant. Thus, physically sensible quantities
depend on the gauge invariant combination $\F-\Upsilon$ which
will be for brevity denoted as the reduced vortex flux in the following.

A solution to the Dirac equation (2.3) with Hamiltonian (2.5) in
background (2.10), that obeys the condition (2.12), can be
presented as
$$<\x|E,n>=\left(\ba{l}
f_n(r,E)e^{i(n+\Upsilon)\varphi}\\[0.2cm]
g_n(r,E)e^{i(n+\Upsilon+s)\varphi}\\ \ea \right), \quad n\in\Z,
\eqno(2.13)$$
where $\Z$ is the set of integer numbers, and the radial functions
$f_n$ and $g_n$ satisfy the system of equations
$$e^{-i\chi_s}[-\partial_r+s(n-\F+\Upsilon)r^{-1}]f_n(r,E)=E g_n(r,E),$$
$$e^{i\chi_s}[\partial_r+s(n-\F+\Upsilon+s)r^{-1}]g_n(r,E)=E f_n(r,E).
\eqno(2.14)$$
When the reduced vortex flux $\F-\Upsilon$ is integer, the requirement of square
integrability for the wave function (2.13) provides its regularity
everywhere on the plane $(x^1,x^2)$, rendering partial Dirac
Hamiltonians for every value of $n$ to be essentially self-adjoint.
When $\F-\Upsilon$ is fractional, the same is valid only for $n\neq
n_0$, where
$$n_0=\io\F-\Upsilon\ic+{1\over2}-{1\over2}s, \eqno(2.15)$$
$\io u\ic$ is the integer part of a quantity $u$ (i.e., the greatest
integer that is less than or equal to $u$). For $n=n_0$, each of the
two linearly independent solutions to system (2.14) meets the
requirement of square integrability. Any particular solution in this
case is characterized by at least one (at most both) of the radial
functions being divergent as $r^{-p}$ ($p<1$) for $r\rightarrow 0$. If
one of the two linearly independent solutions is chosen to have a
regular upper and an irregular lower component, then the other one has
a regular lower and an irregular upper component. Therefore, contrary to
the case of $n\neq n_0$, the partial  Dirac Hamiltonian in the case of
$n=n_0$ is not essentially self-adjoint\footnote{A corollary of the
theorem proven in Ref.\cite{Wei} states that, for the partial Dirac
Hamiltonian to be essentially self-adjoint, it is necessary and
sufficient that a non-square-integrable solution exist.}. The 
Weyl - von Neumann theory of self-adjoint operators (see, e.g., 
Refs.\cite{Akhie,Reed}) has to be employed in order to consider the
possibility of a self-adjoint extension in the case of $n=n_0$. It can
be shown that the self-adjoint extension exists indeed and is
parametrized by one continuous real variable denoted in the following
by $\Theta$. Thus, the partial Dirac Hamiltonian in the case of $n=n_0$
is defined on the domain of functions obeying the condition

$$\cos\bigl(s{\Theta\over2}+{\pi\over4}\bigr)\lim_{r\rightarrow 0}(\mu
r)^Ff_{n_0}=-e^{i\chi_s}\sin\bigl(s{\Theta\over2}+{\pi\over4}\bigr)
\lim_{r\rightarrow 0}(\mu r)^{1-F}g_{n_0}, \eqno(2.16)$$
where $\mu>0$ is the parameter of the dimension of inverse length and
$$
F=s\fo\F-\Upsilon\fc+{1\over2}-{1\over2}s,
\eqno(2.17)$$
$\fo u\fc=u-\io u\ic$ is the fractional part of a quantity $u$,
$0\leq\fo u\fc<1$; note here that Eq.(2.16) implies that $0<F<1$,
since, for $F={1\over2}-{1\over2}s$, both $f_{n_0}$ and
$g_{n_0}$ obey the condition of regularity at $r=0$. Note also that Eq.(2.16) 
is periodic in $\Theta$ with the period of $2\pi$; therefore,
without a loss of generality, all permissible values of $\Theta$ will be
restricted in the following to the range $-\pi\leq\Theta\leq\pi$.

All solutions to the Dirac equation in the background of a singular
magnetic vortex correspond to the continuous spectrum and, therefore,
obey the orthonormality condition
$$
\int d^2x<E,n|\x><\x|E',n'>={\delta(E-E')\over\sqrt{|EE'|}}\delta_{nn'}.
\eqno(2.18)$$
In the case of $0<F<1$ one can get the following expressions
corresponding to the regular solutions with $sn>sn_0$:
$$
\left(\ba{c} f_n\\ g_n \\ \ea \right) ={1\over2\sqrt{\pi}}
\left(\ba{c}
J_{l-F}(kr)e^{i\chi_s}\\[0.2cm]
\sgn(E)J_{l+1-F}(kr)\\ \ea \right), \qquad l=s(n-n_0), \eqno(2.19)$$
the regular solutions with $sn<sn_0$:
$$
\left(\ba{c} f_n\\ g_n \\ \ea \right)
={1\over2\sqrt{\pi}}
\left(\ba{c}
J_{l'+F}(kr)e^{i\chi_s}\\[0.2cm]
-\sgn(E)J_{l'-1+F}(kr) \ea \right), \qquad l'=s(n_0-n), \eqno(2.20)$$
and the irregular solution:
$$
\left(\ba{c} f_{n_0}\\ g_{n_0} \\ \ea \right)
={1\over 2\sqrt{\pi[1+\sin(2\nu_E)\cos(F\pi)]}}\times$$
$$\times \left(\ba{c}
[\sin(\nu_E)J_{-F}(kr)+\cos(\nu_E)J_F(kr)]e^{i\chi_s}\\[0.2cm]
\sgn(E)[\sin(\nu_E)J_{1-F}(kr)-\cos(\nu_E)J_{-1+F}(kr)]\\ \ea \right);
\eqno(2.21)$$
here $k=|E|$, $J_{\rho}(u)$ is the Bessel function of order
$\rho$ and
$$
\sgn(u)=\left\{\ba{cc}
1,& u>0\\
-1,& u<0\\ \ea \right\}.
$$
Substituting the asymptotic form of Eq.(2.21) at $r\rightarrow 0$
into Eq.(2.16), one arrives at the relation between the parameters
$\nu_E$ and $\Theta$:
$$
\tan(\nu_E)=\sgn(E)\bigl({k\over2\mu}\bigr)^{2F-1}\,
{\Gamma(1-F)\over\Gamma(F)}\tan\bigl(s{\Theta\over2}+{\pi\over4}\bigr),
\eqno(2.22)$$
where $\Gamma(u)$ is the Euler gamma function.

Using an explicit form of solutions (2.19) -- (2.21), all vacuum
polarization effects can be determined.

\section{Fermion Number}

In the second-quantized theory in 2+1-dimensional space-time the
operator of the fermion number is given by the expression
$$
\hat{\N}=\int d^2x{1\over2}[\Psi^+(\x,t),\Psi(\x,t)]_-=\si
[a_\lambda^+a_\lambda-b_\lambda^+b_\lambda-{1\over2} \sgn(E_\lambda)],
\eqno(3.1)$$
and, consequently, its vacuum expectation value takes the form
$$
\N\equiv<\vac|\hat{\N}|\vac>=-{1\over2}\si
\sgn(E_\lambda)=-{1\over2}\int d^2x\, \tr<\x|\sgn(H)|\x>. \eqno(3.2)$$
From general arguments, one could expect that the last quantity
vanishes due to cancellation between the contributions of positive and
negative energy solutions to the Dirac equation (2.3). Namely this
happens in a lot of cases. That is why every case of a nonvanishing value
of $\N$ deserves a special attention.

Considering the case of the background in the form of a singular
magnetic vortex (2.10) -- (2.11), one can notice that the contribution
of the regular solutions (2.19) and (2.20) is cancelled upon summation
over the sign of energy, whereas the irregular solution (2.21) yields a
nonvanishing contribution to $\N$ (3.2). Defining the vacuum fermion
number density
$$ \N_{\x}=-{1\over2}\tr<\x|\sgn(H)|\x>, \eqno(3.3.)$$
we get
$$
\N_\x=-{1\over8\pi}\int\limits_0^\infty dkk\biggl\{
A\biggl({k\over\mu}\biggr)^{2F-1}
\biggl[\Lp+\Lm\biggr]\biggl[J_{-F}^2(kr)+$$
$$+J_{1-F}^2(kr)\biggr]+2\biggl[\Lp-\Lm\biggr]\biggl[J_{-F}(kr)J_F(kr)-
J_{1-F}(kr)J_{-1+F}(kr)\biggr]+$$
$$+A^{-1}\biggl({k\over\mu}\biggr)^{1-2F}\biggl[\Lp+\Lm\biggl]
\biggr[J_F^2(kr)+J_{-1+F}^2(kr)\biggr]\biggr\}, \eqno(3.4)$$
where
$$A=2^{1-2F}{\Gamma(1-F)\over\Gamma(F)}\tan\left(s{\Theta\over2}+
{\pi\over4}\right), \eqno(3.5)$$
$$L_{(\pm)}=2^{-1}\bigl\{\cos(F\pi)\pm\cosh\bigl[(2F-1)\ln({k\over\mu})+
\ln A\bigr]\bigr\}^{-1}. \eqno(3.6)$$
Transforming the integral in Eq.(3.4), we get the final expression
$$\N_\x=-{\sin(F\pi)\over2\pi^3r^2}\int\limits_0^\infty dw\,
w {K_F^2(w)-K_{1-F}^2(w)\over \cosh[(2F-1)\ln({w\over\mu r})+\ln A]},
\eqno(3.7)$$
where $K_{\rho}(w)$ is the Macdonald function of order $\rho$.
The vacuum fermion number density (3.7) vanishes at half integer values
of the reduced vortex flux ($F={1\over2}$) as well as at $\cos\Theta=0$.
Otherwise, at large distances from the vortex we get
$$
\N_\x{}_{\stackrel{\ds =}{r\rightarrow
\infty}}-(F-{1\over2}){\sin(F\pi)\over2\pi^2r^2}
\left\{\ba{cc}
(\mu r)^{2F-1}A^{-1}{\ds\Gamma({3\over2}-F)\Gamma({3\over2}-2F)\over\ds
\Gamma(2-F)},& 0<F<{1\over2}\\[0.2cm]
(\mu r)^{1-2F}A{\ds\Gamma(F+{1\over2})\Gamma(2F-{1\over2})\over\ds
\Gamma(1+F)},& {1\over2}<F<1 \\ \ea \right. \,.\eqno(3.8)$$

Integrating Eq. (3.7) over the plane ($x^1,x^2$), we obtain the total
vacuum fermion number
$$\N=-{1\over2}\sgn_0\left[(F-{1\over2})\cos\Theta\right], \eqno(3.9)$$
where
$$\sgn_0(u)=\left\{ \ba{cc}
\sgn(u),& u\neq0\\
0,& u=0\\ \ea \right\}.
$$

\section{Current}

Let us regard the 2-dimensional space $(x^1,x^2)$ as a
1+1-dimensional space-time with the Wick-rotated time axis. The
Clifford algebra in this space-time has the exact irreducible
representation with the above $\alpha$ matrices playing now the role of
the $\gamma$ matrices and the $\beta$ matrix playing the role similar
to that of the $\gamma^5$ matrix in a 3+1-dimensional space-time. Introducing
the mass parameter $M$ to tame the infrared divergence, one can define
the trace of two-point causal Green's function with the $\alpha$
matrix inserted between the field operators:
$$
\J(\x,\x'|M)=<\vac|T\Psi^+(\x',0)\ab\Psi(\x,0)|\vac>=$$
$$\qquad\qquad=\tr<\x|\ab(H-iM)^{-1}|\x'>, \eqno(4.1)$$
where $T$ is the symbol of time ordering in 1+1-dimensional space-time.
Inserting the damping factor $(E^2+M^2)^{-z}$ (where $\Re\, z>0$) into
the integral corresponding to Eq.(4.1), we define the matrix element
$$
\J(\x,\x';z|M)=\tr<\x|\ab(H+iM)(H^2+M^2)^{-1-z}|\x'>. \eqno(4.2)$$
Returning to the massless theory in a 2+1-dimensional space-time
($x^1,x^2,t)$, let us define the vacuum current in the conventional way
(compare with Eqs.(3.1) -- (3.3))
$$\j(\x)=<\vac|{1\over2}\left[\Psi^+(\x,t),\ab\,\Psi(\x,t)\right]_-|\vac>=
-{1\over2}\tr<\x|\ab\,\sgn(H)|\x>. \eqno(4.3)$$
One can notice the relation
$$\j(\x)=-{1\over2}\J(\x,\x;-{1\over2}|0), \eqno(4.4)$$
so that in the following the matrix element (4.2) in the coincidence
limit $\x'=\x$ will be regarded as a generalized current.

In the background of a singular magnetic vortex (2.10) -- (2.11) the
radial component
$$
J_r(\x,\x;z|M)=r^{-1}[x^1J_1(\x,\x;z|M)+x^2J_2(\x,\x;z|M)] \eqno(4.5)$$
vanishes, whereas the angular component
$$J_\varphi(\x,\x;z|M)=r^{-1}[x^1J_2(\x,\x;z;|M)-x^2J_1(\x,\x;z|M)]
\eqno(4.6)$$
is nonvanishing. The contribution of the regular solutions (2.19) and
(2.20) to Eq.(4.6) is given by the expression
$$[J_\varphi(\x,\x;z|M)]_\REG={s\over\pi}\int\limits_0^\infty
dk{k^2\over (k^2+M^2)^{1+z}} \biggl[\sum_{l=1}^\infty
J_{l-F}(kr)J_{l+1-F}(kr)-$$ $$\qquad\qquad -\sum_{l'=1}^\infty
J_{l'+F}(kr)J_{l'-1+F}(kr)\biggr].  \eqno(4.7)$$
Performing the summation over $l$ and $l'$, we get
$$[J_\varphi(\x,\x;z|M)]_\REG= {s\over\pi}\int\limits_0^\infty dk
{k^2\over(k^2+M^2)^{1+z}}\biggl\{ FJ_F(kr)J_{-1+F}(kr)-$$
$$-(1-F)J_{1-F}(kr)J_{-F}(kr)-{1\over2}kr
[J_F^2(kr)+J_{-1+F}^2(kr)-J_{-F}^2(kr)-J_{1-F}^2(kr)]\biggr\}.
\eqno(4.8)$$
Transforming the integral in the last expression, we get
$$
J_\varphi(\x,\x;z|M)]_\REG={s\sin(z\pi)\over\pi^2}r^{-1+2z}
\int\limits_{|M|r}^\infty dw{w^2\over(w^2-M^2r^2)^{1+z}}\times$$
$$\times\biggl\{
I_F(w)K_{1-F}(w)-I_{1-F}(w)K_F(w)+{2\sin(F\pi)\over\pi}
[wK_F^2(w)-wK_{1-F}^2(w)-$$
$$-(2F-1)K_F(w)K_{1-F}(w)]\biggr\}, \eqno(4.9)$$
where $I_{\rho}(w)$ is the modified Bessel function of order $\rho$.
The contribution of the irregular solution (2.21) to Eq.(4.6) is given
by the expression
$$
[J_\varphi(\x,\x;z|M)]_\IRREG={s\over2\pi}\int\limits_0^\infty
dk{k\over(k^2+M^2)^{1+z}}\times$$
$$\times \biggl\{
A\biggl({k\over\mu}\biggr)^{2F-1}[(k+iM)\Lp-(k-iM)\Lm]J_{-F}(kr)J_{1-F}(kr)+$$
$$+ [(k+iM)\Lp+(k-iM)\Lm][J_F(kr)J_{1-F}(kr)-
J_{-F}(kr)J_{-1+F}(kr)]-$$
$$-A^{-1}\biggl({k\over\mu}\biggr)^{1-2F}[(k+iM)\Lp-(k-iM)\Lm]J_F(kr)J_{-1+F}(kr)\biggr\},
\eqno(4.10)$$
where $A$ and $L_{(\pm)}$ are given by Eqs. (3.5) and (3.6),
respectively. Transforming the integral in Eq. (4.10), we get
$$
[J_\varphi(\x,\x;z|M)]_\IRREG={s\sin(z\pi)\over\pi^2}r^{-1+2z}
\int\limits_{|M|r}^\infty dw{w^2\over(w^2-M^2r^2)^{1+z}}\times$$
$$\times \biggl(I_{1-F}(w)K_F(w)-I_F(w)K_{1-F}(w)- {2\sin(F\pi)\over\pi}
K_F(w)K_{1-F}(w)\times$$
$$\times \bigl\{\tanh[(2F-1)\ln({w\over\mu r})+\ln A]- {iMr\over
w\cosh[(2F-1)\ln({w\over\mu r})+\ln A]}\bigr\}\biggr). \eqno(4.11)$$
Summing Eqs. (4.9) and (4.11), we get
$$
J_\varphi(\x,\x;z|M)={2s\sin(F\pi)\over\pi^3}\sin(z\pi)r^{-1+2z}
\int\limits_{|M|r}^\infty dw\,{w^2\over(w^2-M^2r^2)^{1+z}}\times$$
$$\times \biggl( w[K_F^2(w)-K_{1-F}^2(w)]-K_F(w)K_{1-F}(w)
\bigl\{2F-1+$$
$$+\tanh[(2F-1)\ln\bigl({w\over\mu r}\bigr)+\ln A]-{iMr\over
w\cosh[(2F-1)\ln({w\over\mu r})+\ln A]}\bigr\}\biggr). \eqno(4.12) $$
Note that Eqs.(4.9) and (4.11) are valid at $-{1\over2}<\Re z<0$, and
their sum, Eq.(4.12), can be continued analytically to the half plane
$\Re z<0$. This may look somewhat embarrassing, since the initial
motivation, as presented above, was to consider the case of $\Re z>0$.
However, the situation can be cured by means of analytic continuation
using partial integration. Namely, integrating Eq. (4.12) by parts, we
get the expression $$J_\varphi(\x,\x;z|M)=-{s\sin(F\pi)\over\pi^3}\,
{\sin(z\pi)\over z}r^{-1+2z}\int\limits_{|M|r}^\infty
dw(w^2-M^2r^2)^{-z}\times$$
$$\times\biggl[w\bigl[K_F^2(w)-K_{1-F}^2(w)\bigr]+\bigl[K_F^2(w)+K_{1-F}^2(w)
\bigr]\bigl\{ w\tanh\bigl[(2F-1)\ln\bigl({w\over\mu r}\bigr)+\ln A
\bigr]-$$
$$-{iMr\over\cosh[(2F-1)\ln({w\over\mu r})+\ln A]}\bigr\}-
{K_F(w)K_{1-F}(w)\over \cosh[(2F-1)\ln({w\over\mu r})+\ln A]}\times$$
$$\times \biggl( {2F-1\over\cosh[(2F-1)\ln({w\over\mu r})+\ln
A]}+{iMr\over w}\bigl\{(2F-1)\tanh[(2F-1)\ln\bigl({w\over\mu r}\bigr)+
\ln A\bigr]+1\bigr\}\biggr)\biggr], \eqno(4.13)$$
which is continued analytically to the domain $\Re z<1$. Integrating
Eq.(4.12) by parts $N$ times, one can get the expression for the
generalized current which is continued analytically to the domain $\Re
z<N$.

Taking into account Eq.(4.4), we get the following expression for the
vacuum current:
$$j_\varphi(\x)={s\sin(F\pi)\over\pi r^2}\biggl\{
{(F-{1\over2})^2\over4\cos(F\pi)}-{1\over\pi^2}\int\limits_0^\infty dw\,
wK_F(w)K_{1-F}(w)\times$$
$$\qquad\qquad\times \tanh\bigl[[(2F-1)\ln\bigl({w\over\mu r}\bigr)
+\ln A\bigr]\bigr\}; \eqno(4.14)$$
recall that the radial component $j_r(\x)$ is vanishing, see Eq. (4.5).
At $\cos\Theta=0$ we get
$$j_\varphi(\x)={s\tan(F\pi)\over4\pi
r^2}(F-{1\over2})(F-{1\over2}\pm1), \qquad \Theta=\pm s{\pi\over2}.
\eqno(4.15)$$
At half integer values of the reduced vortex flux $(F={1\over2})$,
taking into account the relation
$$A|_{F={1\over2}}=\tan\bigl(s{\Theta\over2}+{\pi\over4}\bigr),
\eqno(4.16)$$
we get
$$j_\varphi(\x)|_{F={1\over2}}=-{\sin\Theta\over4\pi^2r^2}.
\eqno(4.17)$$
If $\cos\Theta\neq0$ and $F\neq{1\over2}$, then at large distances from
the vortex we get $$j_\varphi(\x){}_{\stackrel{\ds =}{r\rightarrow
\infty}}\, {s\tan(F\pi)\over4\pi r^2}|F-{1\over2}|(|F-{1\over2}|-1).
\eqno(4.18)$$

\section{Parity Breaking Condensate}

Since the twodimensional massless Dirac Hamiltonian (2.5) anticommutes
with the $\beta$ matrix
$$[H,\beta]_+=0, \eqno(5.1)$$
the Dirac equation (2.3) is invariant under the parity transformation
$$E_\lambda\rightarrow -E_\lambda, \qquad <\x|\lambda>\rightarrow
\beta<\x|\lambda>. \eqno(5.2)$$
However, this invariance is violated by the boundary condition (2.16), 
unless $\cos\Theta=0$. Consequently, the parity breaking
condensate emerges in the vacuum:
$$\C_\x=<\vac|{1\over2}[\Psi^+(\x,t),\beta\,\Psi(\x,t)]_-|\vac>=$$
$$\qquad =-{1\over2}\tr<\x|\beta\,\sgn(H)|\x>. \eqno(5.3)$$

Let us start with the regularized condensate
$$\C_\x(z|M)=-{1\over2}\tr<\x|\beta\, H(H^2+M^2)^{-{1\over2}-z}|\x>.
\eqno(5.4)$$
The contribution of the regular solutions (2.19) and (2.20) to Eq.
(5.4) is cancelled upon summation over the sign of energy. Thus, only
the contribution of the irregular solution (2.21) to Eq. (5.4) survives:
$$
\C_\x(z|M)=-{1\over8\pi}\int\limits_0^\infty {dk\,
k^2\over(k^2+M^2)^{{1\over2}+z}}\biggl\{
A\biggl({k\over\mu}\biggr)^{2F-1}\bigl[\Lp+\Lm\bigr]
\bigl[J_{-F}^2(kr)-$$
$$-J_{1-F}^2(kr)\bigr]+2\bigl[\Lp-\Lm\bigr]
\bigl[J_{-F}(kr)J_F(kr)+J_{1-F}(kr)J_{-1+F}(kr)\bigr]+$$
$$+A^{-1}\biggl({k\over\mu}\biggr)^{1-2F}\bigl[\Lp+\Lm\bigr]
\bigl[J_F^2(kr)-J_{-1+F}^2(kr)\bigr]\biggr\}, \eqno(5.5)$$
where $A$ and $L_{(\pm)}$ are given by Eqs.(3.5) and (3.6),
respectively. Transforming the integral in Eq.(5.5), we get
$$
\C_\x(z|M)=-{\sin(F\pi)\over2\pi^3}\cos(z\pi)r^{2(z-1)}
\int\limits_{|M|r}^\infty dw\, w^2(w^2-M^2r^2)^{-{1\over2}-z}\times$$
$$\qquad \times {K_F^2(w)+K_{1-F}^2(w)\over \cosh[(2F-1)\ln({w\over\mu
r})+\ln A]}, \eqno(5.6)$$
where $\Re z<{1\over2}$. Then the vacuum condensate (5.3) is given by
the following expression:
$$\C_\x\equiv \C_\x(0|0)=- {\sin(F\pi)\over2\pi^3r^2}\int\limits_0^\infty dw\,
w{K_F^2(w)+K_{1-F}^2(w)\over\cosh[(2F-1)\ln({w\over\mu r})+\ln A]}.
\eqno(5.7)$$
Evidently, Eq.(5.7) vanishes if $\cos\Theta=0$. At half integer values
of the reduced vortex flux $(F={1\over2})$, we get
$$\C_\x\big|_{F={1\over2}}=-{\cos\Theta\over 4\pi^2r^2}. \eqno(5.8)$$
At large distances from the vortex we get
$$
\C_\x{}_{\stackrel{\ds =}{r\rightarrow
\infty}}-{\sin(F\pi)\over2\pi^2r^2} \left\{\ba{cc} (\mu
r)^{2F-1}A^{-1}{\ds\Gamma({3\over2}-F)\Gamma({3\over2}-2F)\over \ds
\Gamma(1-F)}, & 0<F<{1\over2}\\[0.2cm]
(\mu r)^{1-2F}A {\ds\Gamma(F+{1\over2})\Gamma(2F-{1\over2})\over\ds
\Gamma(F)},& {1\over2}<F<1\\ \ea \right. \,.\eqno(5.9)$$
Integrating Eq. (5.7) over the plane ($x^1,x^2$), we obtain the total
vacuum condensate
$$
\C\equiv\int d^2x\,\C_\x=-{\sgn_0(\cos\Theta)\over4|F-{1\over2}|}.
\eqno(5.10)$$
Thus, the total vacuum condensate is infinite at $F={1\over2}$ if
$\cos\Theta\neq0$.

\section{Absence of Parity Anomaly}

Let us define the generalized twodimensional axial current
$$\J^3(\x,\x;z|M)=i\tr<\x|\ab\beta(H+iM)(H^2+M^2)^{-1-z}|\x>.
\eqno(6.1)$$
Owing to the twodimensionality, the components of the current (6.1) are
related to the components of the generalized current considered in
Section 4 (Eq.(4.2) at $\x'=\x$)
$$J_r^3=sJ_\varphi, \qquad J^3_\varphi=-sJ_r. \eqno(6.2)$$
In the background of a singular magnetic vortex (2.10) -- (2.11), the
divergence of the current $\J$ is vanishing (see Eq.(4.12))
$$\bpar\cdot\J(\x,\x;z|M)=0, \eqno(6.3)$$
whereas the divergence of the current $\J^3$ is nonvanishing, owing to
the relation
$$\bpar\cdot\J^3(\x,\x;z|M)=s\,\bpar\times\J(\x,\x;z|M). \eqno(6.4)$$

Let us consider the effective action in the Euclidean 1+1-dimensional
space-time
$$S^{\rm eff}_{(1+1)}[\V(\x)]=-\int d^2x\,
\tr<\x|\ln(H\tilde{M}^{-1})|\x>, \eqno(6.5)$$
where $\tilde{M}$ is the parameter of the dimension of mass. The invariance
of action (6.5) under the gauge transformation,
$$
V(\x)\rightarrow \V(\x)+\bpar\,\Lambda(\x),$$
$$<\x|\rightarrow e^{i\Lambda(\x)}<\x|, \qquad |\x>\rightarrow
e^{-i\Lambda(\x)}|\x>, \eqno(6.6)$$
is stipulated by the conservation law
$$\lim_{M\rightarrow 0\atop z\rightarrow 0}\bpar\cdot\J(\x,\x;z|M)=0.
\eqno(6.7)$$
If the conservation law
$$\lim_{M\rightarrow 0\atop z\rightarrow 0}\bpar\cdot\J^3(\x,\x;z|M)=0
\eqno(6.8)$$
holds, then action (6.5) is invariant under the localized version
of the parity transformation (compare with Eq.(5.2))
$$
\V(\x)\rightarrow \V(\x)+\bpar\,\beta\Lambda(\x),$$
$$<\x|\rightarrow e^{i\beta\Lambda(\x)}<\x|, \qquad |\x>\rightarrow
|\x>e^{i\beta\Lambda(\x)}. \eqno(6.9)$$
The breakdown of the latter symmetry,
$$\lim_{M\rightarrow 0\atop z\rightarrow
0}\bpar\cdot\J^3(\x,\x;z|M)\neq0, \eqno(6.10)$$
is denoted as a parity anomaly, i.e., the axial anomaly in
1+1-dimensional space-time.

Note that in classical theory both gauge and localized parity
symmetries are conserved. This is reflected by the formal invariance of
action (6.5) under both transformations (6.6) and (6.9). However, in
quantum theory, in order to calculate the effective action in a certain
background, one has to use regularization which in fact breaks the
localized parity symmetry. Namely, one has to substitute
$\ln[(H-iM)\tilde{M}^{-1}]$ for $\ln(H\tilde{M}^{-1})$ into Eq. (6.5),
where the regulator mass $M$ is the symmetry breaking parameter. That
is why the generalized currents (4.2) at $\x'=\x$ and (6.1) come into
play. The divergence of the latter current can be presented in the form
$$
\bpar\cdot\J^3(\x,\x;z|M)=
2\tilde{\zeta}_\x(z|M)-2M^2\tilde{\zeta}_\x(z+1|M)-$$
$$\qquad\qquad -4iM\C_\x(z+{1\over2}|M), \eqno(6.11)$$
where $\C_\x(z|M)$ is the regularized condensate (5.4) and
$$\tilde{\zeta}_\x(z|M)=\tr<\x|\beta(H^2+M^2)^{-z}|\x> \eqno(6.12)$$
is the modified (by insertion of the $\beta$ matrix) zeta function
density.

In the background of a singular magnetic vortex the conservation law
(6.7) holds, as a consequence of Eq.(6.3). Incidentally, the
regularized condensate is given by Eq.(5.6). As to the
modified zeta function density, the following expression can be
obtained similarly to the above:
$$
\tilde{\zeta}_\x(z|M)={\sin(F\pi)\over\pi^3}
\sin(z\pi)r^{2(z-1)}\int\limits_{|M|r}^\infty dw\, w(w^2-M^2r^2)^{-z}\times$$
$$\times \biggl\{K_F^2(w)-K_{1-F}^2(w)+
\bigl[K_F^2(w)+K_{1-F}^2(w)\bigr]\tanh\bigl[(2F-1)\ln\bigl({w\over \mu
r}\bigr)+\ln A\bigr]\biggr\}, \eqno(6.13)$$
where $\Re z<1$. Extending the domain of definition in $z$ in Eqs.(5.6) 
and (6.13) by means of integration by parts, we get
$$
\C_\x(z|M)=-{\sin(F\pi)\over\pi^3}\,
{\cos(z\pi)\over1-2z}r^{2(z-1)}\int\limits_{|M|r}^\infty\,
{dw(w^2-M^2r^2)^{{1\over2}-z}\over \cosh[(2F-1)\ln({w\over\mu r})+\ln
A]}\times$$
$$\times\biggl\{
\bigl(F-{1\over2}\bigr)\bigl[K_F^2(w)-K_{1-F}^2(w)\bigr]
+2wK_F(w)K_{1-F}(w)+$$
$$+\bigl(F-{1\over2}\bigr)\bigl[K_F^2(w)+K_{1-F}^2(w)\bigr]
\tanh\bigl[(2F-1)\ln\bigl({w\over\mu r}\bigr)+\ln A\bigr]\biggr\},
\eqno(6.14)$$
where $\Re z<{3\over2}$ and
$$
\tilde{\zeta}_\x(z|M)={\sin(F\pi)\over\pi^3}\, {\sin(z\pi)\over1-z}
r^{2(z-1)}\int\limits_{|M|r}^\infty\, {dw\over w}(w^2-M^2r^2)^{1-z}\times$$
$$\times \biggl\{ {1\over2}\bigl[K_F^2(w)-K_{1-F}^2(w)\bigr]+
\bigl[FK_F^2(w)+(1-F)K_{1-F}^2+2w K_F(w)K_{1-F}(w)\bigr]\times$$
$$\times \tanh\bigl[(2F-1)\ln\bigl({w\over\mu r}\bigr)+\ln A\bigr]+
\bigl(F-{1\over2}\bigr)\bigl[K_F^2(w)+K_{1-F}^2(w)\bigr]
\tanh^2\bigl[(2F-1)\ln\bigl({w\over\mu r}\bigr)+\ln A\bigr]\biggl\},
\eqno(6.15)$$
where $\Re z<2$. Using the last representations, we get
$$
\lim_{M\rightarrow 0}M\C_\x(z+{1\over2}|M)=\lim_{M\rightarrow
0}M^2\tilde{\zeta}_\x(z+1|M)=0, \quad \Re z<1, \eqno(6.16)$$
and, consequently,
$$
\lim_{M\rightarrow 0}\bpar\cdot\J^3(\x,\x;z|M)=2\tilde{\zeta}_\x(z|0),
\quad \Re z<1. \eqno(6.17)$$
One can easily get
$$
\tilde{\zeta}_\x(z|0)={\sin(F\pi)\over\pi^3}\sin(z\pi)r^{2(z-1)}\biggl\{
{\sqrt{\pi}\over2}\, {\Gamma(1-z)\over\Gamma({3\over2}-z)}
\bigl(F-{1\over2})\Gamma(F-z)\Gamma(1-F-z)+$$
$$+\int\limits_0^\infty dw\, w^{1-2z}\bigl[K_F^2(w)+K_{1-F}^2(w)\bigr]
\tanh\bigl[(2F-1)\ln\bigl({w\over\mu r}\bigr)+\ln A\bigr]\biggr\};
\eqno(6.18)$$
in particular, at half integer values of the reduced vortex flux:
$$
\tilde{\zeta}_\x(z|0)\big|_{F={1\over2}}={s\sin\Theta\over
2\pi^{\sss3\over\sss2}}\, {\Gamma({1\over2}-z)\over\Gamma(z)} r^{2(z-1)};
\eqno(6.19)$$
and for $\cos\Theta=0$:
$$
\tilde{\zeta}_\x(z|0)=\pm{\sin(F\pi)\over2\pi^{\sss3\over\sss2}}\,
{\Gamma({3\over2}-z\pm F\mp{1\over2})\Gamma({1\over2}-z\mp
F\pm{1\over2})\over \Gamma(z)\Gamma({3\over2}-z)}\, r^{2(z-1)}, \quad
\Theta=\pm s{\pi\over2}. \eqno(6.20)$$
Consequently, we obtain
$$\tilde{\zeta}_\x(0|0)=0, \qquad \x\neq0. \eqno(6.21)$$

Thus, the anomaly is absent everywhere on the plane with the puncture at
$\x=0$. This looks rather natural, since the twodimensional anomaly
density $2\tilde{\zeta}_\x(0|0)$ is usually identified with the
quantity ${1\over\pi}\bpar\times\V(\x)$ \cite{Schw, Shei, Jack}, and
the last quantity in the present case vanishes everywhere on the
punctured plane, see Eq.(2.11). We see that the natural expectations
are confirmed, provided that the boundary conditions at the puncture
are chosen to be physically acceptable, i.e., compatible with the
self-adjointness of the Hamiltonian\footnote{The opposite claim of the
authors of Ref. \cite{Sold} is not justified.}; we conclude that the
leakage of the anomaly, unlike that of the vacuum condensate or of the
vacuum fermion number, does not happen.

We might finish here the discussion of the anomaly problem in the
background of a singular magnetic vortex. However, there remains a
purely academic question: what is the anomaly density in 
background (2.10) -- (2.11) on the whole plane (without puncturing
$\x=0$)? Just due to a confusion persisting in the literature
\cite{Sold,Mor}, we shall waste now some time to clarify this, otherwise
inessential, point.

The background field strength (2.11), when considered on the whole
plane, is interpreted in the sense of a distribution (generalized
function), i.e., a functional on a set of suitable test functions
$f(\x)$:
$$\int d^2x\,{1\over\pi}\bpar\times\V(\x)f(\x)=2\F f(0); \eqno(6.22)$$
here $f(\x)$ is a continuous function. In particular, choosing
$f(\x)=1$, one gets
$$\int d^2x\,{1\over\pi}\bpar\times\V(\x)=2\F. \eqno(6.23)$$
Considering the anomaly density on the whole plane, one is led to study
different limiting procedures as $r\rightarrow 0$ and $z\rightarrow 0$
in Eq.(6.18). So, the notorious question is, whether the anomaly
density $2\tilde{\zeta}_\x$ can be interpreted in the sense of a
distribution which coincides with the distribution
${1\over\pi}\bpar\times\V(\x)$? The answer is resolutely negative, and
this will be immediately demonstrated below.

First, using the explicit form (6.18), we get
$$
\i d^2x\,2\tilde{\zeta}_\x(z|0)=\left\{\ba{cc}
\infty,&z\neq0\\ 0,&z=0\\ \ea \right.\, ;\eqno(6.24)$$
therefore, the anomaly functional cannot be defined on the same set of
test functions as that used in Eq.(6.22) (for example, the test
functions have to decrease rapidly enough at large (small) distances in
the case of $z>0$ ($z<0$)). Moreover, if one neglects the requirement
of self-consistency, permitting a different set of test functions for
the anomaly functional, then even this will not save the situation. Let
us use the test functions which depend on $z$ and are adjusted in such
a way that the quantity
$$
\A=\lim_{z\rightarrow 0}\i d^2x\,2\tilde{\zeta}_\x(z|0)f(\x;z)\eqno(6.25)$$
is finite. Certainly, this quantity can take values in a rather wide
range, but it cannot be made equal to the right-hand side of Eq.
(6.23). Really, the only source of the dependence on $\F$ in the
integral in Eq. (6.25) is the factor $\tilde{\zeta}_\x(z|0)$, and the
latter, as is evident from Eq. (6.18), depends rather on $\fo\F\fc$,
than on $\F$ itself, thus forbidding the linear dependence of $\A$ on
$\F$. In particular, let us choose the test function $f(\x;z)$ with the
asymptotics at small distances:
$$
f(\x;z){}_{\stackrel{\ds=}{r\rightarrow 0}} [1+(\tilde{\mu}r)^{-2}]^{z-1}
\tilde{\mu}^{2z}, \eqno(6.26)$$
where $\tilde{\mu}$ is the parameter of dimension of mass, and the
asymptotics at large distances providing the vanishing of the integral
in Eq.(6.25) at the upper limit. Choosing the case of $\cos\Theta=0$
for simplicity and taking into account Eq. (6.20), we get
$$
\A=-2(F-{1\over2}\pm{1\over2}), \qquad \Theta=\pm
s{\pi\over2}, \eqno(6.27)$$
which differs clearly from $2\F$.

Thus, in a singular background the conventional relation between the
anomaly density and the background field strength is valid only in
the space with punctured singularities. If the singularities are not
punctured, then the anomaly density and the background field strength
can be interpreted in the sense of distributions, but, contrary to the
assertion of the authors of Refs.\cite{Sold,Mor}, the conventional
relation is not valid.

\section{Angular Momentum}

Let $\hat{M}$ be an operator in the first-quantized theory, which
commutes with the Dirac Hamiltonian
$$
[\hat{M},H]_-=0. \eqno(7.1)$$
Then, in the second-quantized theory, the vacuum expectation value of the
dynamical variable corresponding to $\hat{M}$ is presented in the form
$$
\M=\i d^2x\,\M_\x, \eqno(7.2)$$
where
$$
\M_\x=<\vac|{1\over2}\bigl[\Psi^+(\x,t),\hat{M}\,\Psi(\x,t)\bigr]_-|\vac>=
$$
$$\quad =-{1\over2}\tr<\x|\hat{M}\,\sgn(H)|\x>. \eqno(7.3)$$
The commutation relation (7.1) is the evidence of invariance of the
theory with $\hat{M}$ being the generator of the symmetry
transformations. Since, in the background of a singular magnetic vortex
(2.10) -- (2.11), there is invariance with respect to  rotations
around the location of the vortex, one can take $\hat{M}$ as the
generator of rotations -- the operator of angular momentum in the
first-quantized theory (see \cite{SitR96} for more details):
$$\hat{M}=-i\x\times\bpar-\Upsilon+{1\over2}s\beta. \eqno(7.4)$$
Note that the eigenvalues of the operator $\hat{M}$ (7.4) on spinor
functions satisfying condition (2.12) are half integer.

Decomposing Eq.(7.4) into the orbital and spin parts, we get in the
second-quantized theory
$$
\M_\x=\L_\x+\S_\x, \eqno(7.5)$$
where
$$
\L_\x={1\over2}\tr<\x|(i\x\times\bpar+\Upsilon)\,\sgn(H)|\x> \eqno(7.6)$$
and
$$
\S_\x=-{1\over4}s\,\tr<\x|\beta\,\sgn(H)|\x>. \eqno(7.7)$$
Since the vacuum spin density (7.7) is related to the vacuum condensate
(5.3),
$$
\S_\x={1\over2}s\,C_\x, \eqno(7.8)$$
there remains only the vacuum orbital angular momentum density (7.6) to
be considered.

Let us start, as in Section 5, with the regularized quantity
$$
\L_\x(z|M)={1\over2}\tr<\x|(i\x\times\bpar+\Upsilon)
H(H^2+M^2)^{-{1\over2}-z}|\x>. \eqno(7.9) $$
The contribution of the regular solutions (2.19) and (2.20) to Eq.(7.9) 
is cancelled upon summation over the sign of energy, whereas the
contribution of the irregular solution (2.21) to Eq. (7.9) survives
$$\L_\x(z|M)=-{1\over8\pi}\int\limits_0^\infty\, {dk\,k^2\over
(k^2+M^2)^{{1\over}+z}} \biggl\{ A\biggl({k\over\mu}\biggr)^{2F-1}\bigl[
\Lp+\Lm\bigr]\bigl[n_0J_{-F}^2(kr)+$$
$$+(n_0+s)J_{1-F}^2(kr)\bigr]+2
\bigl[\Lp-\Lm\bigr]\bigl[n_0J_{-F}(kr)J_F(kr)-$$
$$-(n_0+s)J_{1-F}(kr)J_{-1+F}(kr)\bigr]+A^{-1}\biggl({k\over\mu}
\biggr)^{1-2F} \bigl[\Lp+\Lm\bigr]\bigl[n_0J_F^2(kr)+$$
$$+(n_0+s)J_{-1+F}^2(kr)\bigr]\biggr\}, \eqno(7.10)$$
where $A$ and $L_{(\pm)}$ are given by Eqs.(3.5) and (3.6).
Transforming the integral in Eq.(7.10), we get
$$\L_\x(z|M)=-{\sin(F\pi)\over2\pi^3}\cos(z\pi)
r^{2(z-1)}\int\limits_{|M|r}^\infty dw\,
w^2(w^2-M^2r^2)^{-{1\over2}-z}\times$$
$$\times {n_0K_F^2(w)-(n_0+s)K_{1-F}^2(w)\over
\cosh[(2F-1)\ln({w\over\mu r})+\ln A]}, \eqno(7.11)$$
where $\Re z<{1\over2}$. Then we get
$$
\L_\x\equiv \L_\x(0|0)=-{\sin(F\pi)\over2\pi^3r^2}\int\limits_0^\infty dw\, w
{n_0K_F^2(w)-(n_0+s)K_{1-F}^2(w)\over \cosh[(2F-1)\ln({w\over\mu
r})+\ln A]}. \eqno(7.12)$$

Summing Eqs.(7.12) and (7.8), taking into account Eqs.(5.7) and
(2.15), we obtain the following expression  for the vacuum angular
momentum density in the background of a singular magnetic vortex
(2.10) -- (2.11):
$$\M_\x=\bigl(\io\F-\Upsilon\ic+{1\over2}\bigr)\,\N_\x, \eqno(7.13)$$
where the vacuum fermion number density $\N_\x$ is given by Eq.(3.7).
Thus, the total vacuum angular momentum takes the form (see Eq.(3.9))
$$\M=-{1\over2}\bigl(\io\F-\Upsilon\ic+{1\over2}\bigr)\,
\sgn_0\bigl[(F-{1\over2})\cos\Theta\bigr]. \eqno(7.14)$$

Concluding this section, let us note that relation (7.1) remains to
be valid if a constant is added to the operator $\hat{M}$. Thus, a
definition which is alternative to Eq.(7.4) has been proposed for the
angular momentum in the first-quantized theory \cite{Wil}:
$$\hat{M}'=-i\x\times\bpar-\F+{1\over2}s\beta. \eqno(7.15)$$
Then, in the second-quantized theory, we get
$$\M'_\x=-s\,\bigl(F-{1\over2}\bigr)\,\N_\x \eqno(7.16)$$
and
$$\M'={1\over2}\,s\,|F-{1\over2}|\,\sgn_0(\cos\Theta).\eqno(7.17)$$
Various arguments pro and contra the physical meaningfulness of the
operator $\hat{M}'$ (7.15) are known in the literature (see Refs.
\cite{Wil, Jack1,Jack2}). However, the crucial point is that the
operator $\hat{M}$ is the generator of rotations, while the operator 
$\hat{M}'$ is
not (the eigenvalues of operator $\hat{M}'$ on spinor functions are not half
integer).

\section{Conclusion}

We have determined the effects of polarization of the massless
fermionic vacuum by a singular magnetic vortex in 2+1-dimensional
space-time. If the quantized massless fermion field belongs to an
irreducible representation of the Clifford algebra, then fermion
number, current, parity breaking condensate, spin and angular momentum
are induced in the vacuum. We have demonstrated that the parity anomaly
is not induced in the vacuum.

All boundary conditions at the location of the vortex provide the Dirac
Hamiltonian to be self-adjoint. The condition $\cos\Theta=0$ is
distinguished, since it corresponds to one of the two components of a
solution to the Dirac equation being regular for all $n$; if
$\Theta=s{\pi\over2}$, then the lower components are  regular, and, if
$\Theta=-s{\pi\over2}$, then the upper components are regular. For this
boundary condition the vacuum current only is nonvanishing, while all
other vacuum polarization effects are vanishing. At half integer values
of the reduced vortex flux the vacuum current and condensate are
nonvanishing, the total condensate being infinite unless
$\cos\Theta=0$, while other vacuum polarization effects are vanishing.
Note also that the vacuum fermion number, spin and angular momentum
change their sign, while the vacuum current and condensate remain unchanged
if one goes over to the inequivalent representation of the Clifford
algebra.

Finally, let us consider the case of a quantized massless fermion
field belonging to the reducible representation composed as a direct
sum of two inequivalent irreducible representations. It follows
immediately from the above that, in the last case, the vacuum current
and condensate only are induced in the vacuum. However, now the vacuum
condensate breaks chiral symmetry rather than parity.  It should be
noted that chiral symmetry breaking in the background of regular
configurations of an external magnetic field has been extensively
discussed in the literature \cite{Gus, Dunn}. One concludes that chiral
symmetry breaking occurs also in the background of a singular
configuration of an external magnetic field, as a result of leakage
from the point of singularity.

The research was supported by the State Foundation for Fundamental
Research of Ukraine (project 2.4/320) and the Swiss National Science
Foundation (grant CEEC/NIS/96-98/7 IP 051219).

\end{document}